\documentclass[traditabstract]{aa}
\usepackage{graphicx}
\usepackage{mathrsfs,amssymb,amsmath}
\usepackage[usenames]{color}
\usepackage{subfigure}

\usepackage[varg]{txfonts}

\newcommand{\bfl}{\begin{flushleft}}
\newcommand{\efl}{\end{flushleft}}

\newcommand{\be}{\begin{equation}}
\newcommand{\ee}{\end{equation}}

\begin{document}

\title{Cosmic-ray acceleration in young protostars}

\author{M. Padovani\inst{1,2}, P. Hennebelle\inst{3}, A. Marcowith\inst{1} \& K. Ferri\`ere\inst{4}}

\authorrunning{M. Padovani et al.}


\institute{Laboratoire Univers et Particules de Montpellier, UMR 5299 du CNRS, Universit\'e de Montpellier, place E. Bataillon, cc072, 34095 Montpellier, France\\
\email{[Marco.Padovani,Alexandre.Marcowith]@umontpellier.fr}
\and
INAF--Osservatorio Astrofisico di Arcetri, Largo E. Fermi 5, 50125 
Firenze, Italy
\and 
CEA, IRFU, SAp, Centre de Saclay, 91191 Gif-Sur-Yvette, France\\
\email{hennebelle@cea.fr}
\and
IRAP, Universit\'e de Toulouse, CNRS, 9 avenue du Colonel Roche, BP 44346, F-31028 Toulouse Cedex 4, France\\
\email{katia.ferriere@irap.omp.eu}
}

\abstract{
The main signature of the interaction between cosmic rays and molecular clouds is the high ionisation degree. 
This decreases towards the densest 
parts of a cloud, where star formation is expected, because of energy losses and magnetic effects. 
However recent observations hint to high 
levels of ionisation in protostellar systems, therefore leading to an apparent contradiction that could be explained by the presence of
energetic particles 
accelerated within young protostars. 
Our modelling consists of a set of conditions that has to be satisfied in order to have an efficient particle acceleration through
the diffusive shock acceleration mechanism. We find that jet shocks can be strong accelerators of protons which can be boosted up to 
relativistic energies. Another possibly efficient acceleration site is located at protostellar surfaces, where shocks caused by impacting
material during the collapse phase are strong enough to accelerate protons. Our results demonstrate the possibility of accelerating particles during the early phase of a proto-Solar-like system and 
can be used as an argument to support available observations. 
The existence of an internal source of energetic particles can have a strong and unforeseen impact on the star and 
planet formation process as well as on the formation of pre-biotic molecules.}
\maketitle

\section{Introduction}
It is largely accepted that Galactic cosmic rays, which pervade the interstellar medium, are likely produced in shock waves in supernova 
remnants
(Drury~\cite{d83}). Cosmic 
rays activate the rich chemistry that is observed in a molecular cloud
(Duley \& Williams~\cite{dw84}) and they also regulate its collapse timescale (Balbus \& Hawley \cite{bh91}; Padovani et al.~\cite{pg14}), determining the efficiency of 
star 
and planet formation, but they cannot penetrate in to the densest parts of a molecular cloud, where the formation of stars is expected, because 
of energy 
losses and magnetic field deflections (Padovani et al.~\cite{pgg09}; Padovani \& Galli~\cite{pg11,pg13}; Padovani et al.~\cite{phg13}; 
Cleeves et al.~\cite{ca13}). Recently, observations towards young protostellar systems showed a surprisingly high value of the 
ionisation rate (Ceccarelli et al.~\cite{cd14}; Podio et al.~\cite{pl14}), the main indicator of the presence of cosmic rays in molecular 
clouds. Synchrotron emission, the typical feature of relativistic 
electrons, was also detected towards the bow shock of a T Tauri star (Ainsworth et al.~\cite{as14}). 
Nevertheless, the origin of these signatures peculiar to energetic particles 
is still puzzling. Here we show that particle acceleration can be driven by shock waves occurring {\em within} protostars. 

\section{Particle acceleration in shocks}\label{conditionsDSA}
This works focuses on shock acceleration by means of the diffusive shock acceleration (DSA) mechanism.
Also known as first-order Fermi acceleration, DSA is a process where 
charged particles systematically gain energy while crossing a shock front. 
Multiple shock crossings allow the particle energy to rapidly increase, reaching the relativistic domain. The motion of particles back and 
forth from upstream to downstream requires the presence of magnetic fluctuations that produce a scattering of the pitch angle, namely the angle 
between the particle's velocity and the mean magnetic field
(Drury~\cite{d83}; Kirk~\cite{k94}). 
We argue below that the accelerated particles themselves can produce the necessary fluctuations to maintain DSA around shocks,
as discussed, e.g., in Bell~(\cite{b78}). 

In the following subsections we describe all the conditions that have to be satisfied in order to effectively accelerate protons and electrons
through DSA. The acceleration of helium and heavier nuclei will be presented in a following paper. 
All the constraints limiting the maximum energy of the accelerated particles are written as functions of the upstream
flow velocity in the shock 
reference frame, $U_{\rm sh}=v_{\rm fl}-v_{\rm sh}$, $v_{\rm fl}$ and $v_{\rm sh}$ are the flow and the shock velocities in the 
observer reference frame, respectively.
Our aim is to compute the maximum energy, $E_{\rm max}$, reached by a particle in the regime where ions and neutrals are coupled.
As explained in Sect.~\ref{ionneutralfriction}, in this case the damping of the Alfv\'en waves, which determines the confinement of particles, is 
weak and particle acceleration is more effective.

\subsection{Condition on shock velocity}\label{csva}
In order to have efficient particle acceleration, the flow has to be supersonic and super-Alfv\'enic. These two conditions are combined into
the 
following relation
\be\label{Uhighercsva}
U_{\rm sh,2}>\max\left\{9\times10^{-2}\left[\gamma_{\rm ad} T_{4}(1+x)\right]^{0.5},2\times10^{-4}n_{6}^{-0.5}B_{-5}\right\}\,,
\ee
where $U_{\rm sh,2}$ has units of $10^{2}$~km~s$^{-1}$,
$\gamma_{\rm ad}$ is the adiabatic index, $T_{4}$ the upstream temperature in $10^{4}$~K, $n_{6}$ the total number density of hydrogen in 
$10^{6}$~cm$^{-3}$, $x$ the ionisation fraction,
and $B_{-5}$ the magnetic field strength in $10^{-5}$~G. 
The two terms in square brackets on the right-hand side of Eq.~(\ref{Uhighercsva}) are the ambient (or upstream) sound speed and
the Alfv\'en speed of the total gas in $10^{2}$~km~s$^{-1}$, respectively.

\subsection{Condition on low-energy particle acceleration: collisional losses}\label{colllosses}
We are interested in the acceleration of low-energy particles ($\lesssim100$~MeV$-$1~GeV), because they are responsible for the bulk of the ionisation. 
We have
to verify that the shock acceleration rate is larger than the collisional loss rate ($t^{-1}_{\rm acc}>t^{-1}_{\rm loss}$).
Following
Drury et al.~(\cite{dd96}), the acceleration rate is given by
\be\label{nuacc}
t^{-1}_{\rm acc}=\frac{3.2\times10^{-8}}{\gamma-1}\frac{k_{\rm u}^{\alpha}(r-1)}{r(1+rk_{\rm d}/k_{\rm u})}%
\tilde\mu^{-1}
U_{\rm sh,2}^{2}B_{-5}~\mathrm{s^{-1}}\,,
\ee
where $\tilde\mu=m/m_{p}$ is the particle mass normalised to the proton mass, $k_{\rm u}$ and $k_{\rm d}$ are the diffusion coefficients 
in the upstream and downstream media, respectively, normalised to the Bohm value for protons
\be 
k_{\rm u,d}^{-\alpha}=\frac{\kappa_{\rm u,d}}{\kappa_{\rm B}}=\frac{3eB}{\gamma\beta^{2}m_{p}c^{3}}\kappa_{\rm u,d}\,,
\ee
with $\gamma$ the Lorentz factor, $\beta=\gamma^{-1}\sqrt{\gamma^{2}-1}$,
%
and $r$ is the shock compression ratio.
For a parallel shock, $\alpha=-1$ and $\kappa_{\rm u}=\kappa_{\rm d}$, while for a perpendicular shock, $\alpha=1$ and 
$\kappa_{\rm u}=r\kappa_{\rm d}$.\footnote{A parallel/perpendicular shock is when the shock normal is parallel/perpendicular,
respectively, to the ambient magnetic field.}
%
%
%
The energy loss rate is given by
\be \label{nuloss}
t^{-1}_{\rm loss}=3.2\times10^{-9}\frac{\beta}{\gamma-1} \tilde\mu^{-1}n_{6}L_{-25}~\mathrm{s^{-1}}\,,
\ee
where $L_{-25}$ is the energy loss function (Padovani et al.~\cite{pgg09}) in units of $10^{-25}$~GeV~cm$^{2}$ which was
extended to lower energies including Coulomb losses (for protons, Mannheim \& Schlickeiser~\cite{ms94},
%
%
and electrons, Swartz et al.~\cite{sn71})
%
%
and synchrotron losses 
(Schlickeiser~\cite{s02}).
The maximum energy of accelerated particles set by energy losses, $E_{\rm loss}$,
is found when $t^{-1}_{\rm acc}=t^{-1}_{\rm loss}$, specifically
\be\label{FEloss}
\beta L_{-25} = 10\frac{k_{\rm u}^{\alpha}(r-1)}{r(1+rk_{\rm d}/k_{\rm u})}%
U_{\rm sh,2}^{2}n_{6}^{-1}B_{-5}\,.
\ee

\subsection{Condition on particle acceleration: ion-neutral friction}\label{ionneutralfriction}

The main limit on the possibility of particle acceleration is given by the presence
of an incomplete ionised medium. In fact, the collision rate between ions and neutrals
can be as high as to decrease the effectiveness of the DSA,
damping the particle's self-generated Alfv\'en waves responsible for the particle scattering back and forth the 
shock.
Ions and neutrals are effectively decoupled if the wave frequency is larger than the ion-neutral collision frequency. 
Following Drury et al.~(\cite{dd96}) and accounting for the fact that particles are not fully relativistic,
we find that the critical energy separating these two regimes, $E_{\rm coup}$, is derived by solving the
following relation
\be\label{Edamp}
\gamma\beta=8.5\times10^{-7}\tilde\mu^{-1}T_{4}^{-0.4}(n_{6}x)^{-1.5}B_{-5}^{2}\,.
\ee
If the particle energy is larger than $E_{\rm coup}$, ions and neutrals are coupled.

The upper cut-off energy due to wave damping, $E_{\rm damp}$, 
is set by requiring that the flux of accelerated particles advected downstream by
the flow is equal to the flux of particles lost upstream because of the lack of waves (due to wave damping) to confine the particles. Following Drury et al.~(\cite{dd96}),
using their exact equation for the wave damping rate,
accounting for departures from fully relativistic behaviour, and assuming $U_{\rm sh}$ to be
much larger than the Alfv\'en speed, $E_{\rm damp}$ follows from
\be\label{Ecut}
\gamma\beta^{2}=8.8\times10^{-5}\tilde\mu^{-1}\Xi%
U_{\rm sh,2}^{3}T_{4}^{-0.4}n_{6}^{-0.5}(1-x)^{-1}B_{-5}^{-4}P^{\prime}_{-2}\,,
\ee
where
\be\label{XI}
\Xi=B_{-5}^{4}+1.4\times10^{12}\tilde\mu^{2}\gamma^{2}\beta^{2}
T_{4}^{0.8}n_{6}^{3}x^{2}\,,
\ee
Both Eq.~(\ref{Edamp}) and Eq.~(\ref{Ecut}) are valid for $T\in[10^{2},10^{5}]$~K.
$P^{\prime}_{-2}$ is the fraction of the shock energy ($m_{p}nU_{\rm sh}^{2}$) going into particle acceleration 
in $10^{-2}$.
$P^{\prime}$ is proportional to the shock efficiency $\eta\in[10^{-6},10^{-3}]$ (Bykov~\cite{by04}), which represents the fraction of 
particles extracted from the thermal plasma and injected into the acceleration process by a shock.
We predict both non-relativistic and mildly relativistic accelerated particles and we checked a posteriori that there is no strong back-reaction.
This means that the upstream medium is not warned by these particles that a shock is coming and we can safely assume that the shock 
and the DSA process are unmodified.
In other words, calculations are carried out in the test-particle limit.

If $E_{\rm damp}>E_{\rm coup}$, then $E_{\rm damp}$ is in the coupled regime, namely 
neutrals coherently move with ions and ion-generated waves are weakly damped.
The last inequality can be written by combining Eqs.~(\ref{Edamp}) and~(\ref{Ecut}) as
\be\label{Rratio}
{\mathscr R}
=\frac{10^{2}}{\beta}%
\Xi%
U_{\rm sh,2}^{3}
n_{6}x^{1.5}(1-x)^{-1}%
B_{-5}^{-6}P^{\prime}_{-2}%
>1\,.
\ee

We consider shocks in three types of environments: in jets as well as in accretion
flows in the collapsing envelopes and on the surfaces of protostars. Using the range of parameters
of Table~\ref{paramspace}, we find that 
$\mathscr{R}\ll1$ in protostellar envelopes (Sect.~\ref{fulfilment}).
This is to say that the following two conditions on shock age and geometry (Sect.~\ref{agegeometry})
are only discussed with reference to shocks in jets and on protostellar surfaces.

\subsection{Conditions due to shock age and geometry}\label{agegeometry}
The maximum energy set by the age of the shock, $E_{\rm age}$, 
is found when the acceleration time, given by the inverse of Eq.~(\ref{nuacc}), is equal to the age of the 
shock. The latter can be assumed of the order of the dynamical time of the
jet ($\gtrsim10^{3}$~yr, de Gouveia Dal Pino~\cite{dg95}) or equal to the accretion time in the case of a surface shock ($\sim10^{5}$~yr,
Masunaga \& Inutsuka~\cite{mi00}). 
%
%
Then, $E_{\rm age}$ is computed from
\be\label{Eage}
\gamma-1=10^{3}\frac{k_{\rm u}^{\alpha}(r-1)}{r(1+rk_{\rm d}/k_{\rm u})}%
\tilde\mu^{-1}%
U_{\rm sh,2}^{2}B_{-5}t_{\rm age,3}\,,
\ee
with $t_{\rm age,3}$ in units of $10^{3}$~yr.

A further constraint is given by the geometry of the shock. In particular, the upstream diffusion length, 
$\lambda_{\rm u}=\kappa_{\rm u}/U_{\rm sh}$, has to be at most a given fraction $\epsilon<1$ of the shock radius, $R_{\rm sh}$; 
besides, in the jet configuration particles may also escape in the transverse direction (the shock can be assumed planar as long as the particle's mean free path around the
shock is smaller than the transverse size of the jet, $R_{\perp}$). 
The maximum energy due to upstream escape losses, $E_{\rm esc,u}$, follows from 
\be\label{FEescu}
\gamma\beta^{2}=4.8 k_{\rm u}^{\alpha}%
\tilde\mu^{-1}%
U_{\rm sh,2}B_{-5}\min\left(\epsilon R_{\rm sh,2},R_{\perp,2}\right)\,,
\ee
where both $R_{\rm sh,2}$ and $R_{\perp,2}$ are in units of $10^{2}$~AU.
In the following we assume $\epsilon=0.1$ (Berezhko et al.~\cite{be96}). 
Since jet shocks have a small transverse dimension, there is a further condition for the escape of
particles downstream: the maximum energy due to downstream escape losses, $E_{\rm esc,d}$, is found when the acceleration time, inverse of Eq.~(\ref{nuacc}), is equal to the downstream 
diffusion time, $t_{\rm diff,d}=R_{\perp}^{2}/(4\kappa_{\rm d})$,\footnote{The factor 4 in the denominator comes from the fact that the diffusion process in the perpendicular direction is in two dimensions.} namely
%
%
%
\be\label{FEescd}
\gamma\beta^{2}(\gamma-1)=5.8\frac{(k_{\rm u}k_{\rm d})^{\alpha}(r-1)}{r(1+rk_{\rm d}/k_{\rm u})}%
\tilde\mu^{-1}%
(U_{\rm sh,2}B_{-5}R_{\perp,2})^{2}\,.
\ee
Finally, if the shock is supersonic and super-Alfv\'enic (Eq.~\ref{Uhighercsva}) and if $\mathscr{R}>1$ (Eq.~\ref{Rratio}),
the maximum energy reached by a particle is $E_{\rm max}=\min[E_{\rm loss},E_{\rm damp},E_{\rm age},E_{\rm esc,u},E_{\rm esc,d}]$.

\section{Potential particle acceleration sites}\label{fulfilment}
In this Section, we identify and characterise possible sites of particle acceleration in protostars.
In particular, we consider accretion flows in the collapsing envelopes and on protostellar surfaces as well as jets.
The required parameters needed to prove the effectiveness of shock acceleration are shown in Table~\ref{paramspace}.

It is straightforward to verify that Eq.~(\ref{Rratio}) 
is not fulfilled in envelopes (${\mathscr R}\ll1$). The ionisation fraction and the
shock velocity are too small, quenching the particle acceleration. 
The magnetic field strength is also as large as to produce a sub-Alfv\'enic shock.
This rules out envelopes as a possible shock acceleration
sites. 
From now on we will focus on shocks in jets and on protostellar surfaces.

\begin{table}[!ht]
\setlength{\tabcolsep}{3pt}
\caption{Values of the parameters described in the text.}
\begin{center}
\begin{tabular}{cccccc}
\hline\hline
site$^{*}$ & $U_{\rm sh}$ & T & $n$ & $x$ & $B$\\
     & $\mathrm{[km~s^{-1}]}$ & $\mathrm{[K]}$ & $\mathrm{[cm^{-3}]}$ & & $\mathrm{[G]}$\\ 
\hline
${\cal E}$ & $1-10$ & $50-100$ & $10^{7}-10^{8}$ & $\lesssim10^{-6}$ & $10^{-3}-10^{-1}$\\
${\cal J}$ & $40-160$ & $10^{4}-10^{5}$ & $10^{4}-10^{7}$ & $0.01-0.9$ & $5\times10^{-5}-10^{-3}$\\
${\cal S}$ & 260 & $9.4\times10^{5}$ & $1.9\times10^{12}$ & $0.01-0.9$ & $10^{-1}-10^{3}$\\
\hline
\end{tabular}\\
$^{*}{\cal E}$ = envelope; ${\cal J}$ = jet; ${\cal S}$ = protostellar surface.
\end{center}
\label{paramspace}
\end{table}
\normalsize

\vspace{-1cm}
\subsection{Jets}\label{jet}
Jets are observed at all stages during the evolution of a protostar (e.g. McCaughrean et al.~\cite{mz02};
Reipurth et al.~\cite{rh97};
Watson \& Stapelfeldt~\cite{ws04}).
Jet speeds, $v_{\rm fl}$, are similar for different classes ($60-300$~km~s$^{-1}$) with shock velocities, $v_{\rm sh}$, of the order
of $20-140$~km~s$^{-1}$ (Raga et al.~\cite{rv02,rn11}; Hartigan \& Morse~\cite{hm07}; Agra-Amboage et al.~\cite{ad11}),
then $U_{\rm sh}=40-160$~km~s$^{-1}$.
The neutral density is between $10^{4}$ and $10^{7}$~cm$^{-3}$ (Lefloch et al.~\cite{lc12}; G\'omez-Ruiz et al.~\cite{gg12})
with temperatures of the order of $10^{4}$~K up to about $10^{6}$~K (Frank et al.~\cite{fr14}).
There is only one theoretical estimate for the magnetic field strength ($B\sim300-500~\mu$G) for Class~II sources
(Te\c sileanu et al.~\cite{tm09,tm12}).
The transverse radius of a jet is about 5~AU, 10~AU,
and 50~AU at 15~AU, 100~AU, and 1000~AU from the source, respectively 
(Cabrit et al.~\cite{cc07}; Hartigan et al.~\cite{he04}).
The ionisation fraction in Class~I and~II are similar, $x\sim0.05-0.9$ (Nisini et al.~\cite{nb05}; Maurri et al.~\cite{mb14}), while
Class~0 jets are mainly molecular (Dionatos et al.~\cite{dn10}).

\subsection{Accretion flows on protostellar surfaces}\label{stellarsurfaceacc}
We use the computational results of the protostellar collapse of an initially homogeneous cloud core described in
Masunaga \& Inutsuka~(\cite{mi00}). 
Their simulation describes the phase of main accretion,
when the protostar mass grows because of the steady accretion from the infalling envelope.
They give the temporal evolution of temperature, density, and flow velocity which,
assuming a stationary shock, is equal
to the shock velocity. 
The radius of the protostar is set to $2\times10^{-2}$~AU and we find that only the last time step of the simulation, 
corresponding to the 
end of the main accretion phase, leads to a strong proton acceleration. Parameters are listed in the third row 
of Table~\ref{paramspace}.


\section{Maximum energy of the accelerated particles}
\begin{figure}[!t]
\begin{center}
\resizebox{\hsize}{!}{\includegraphics{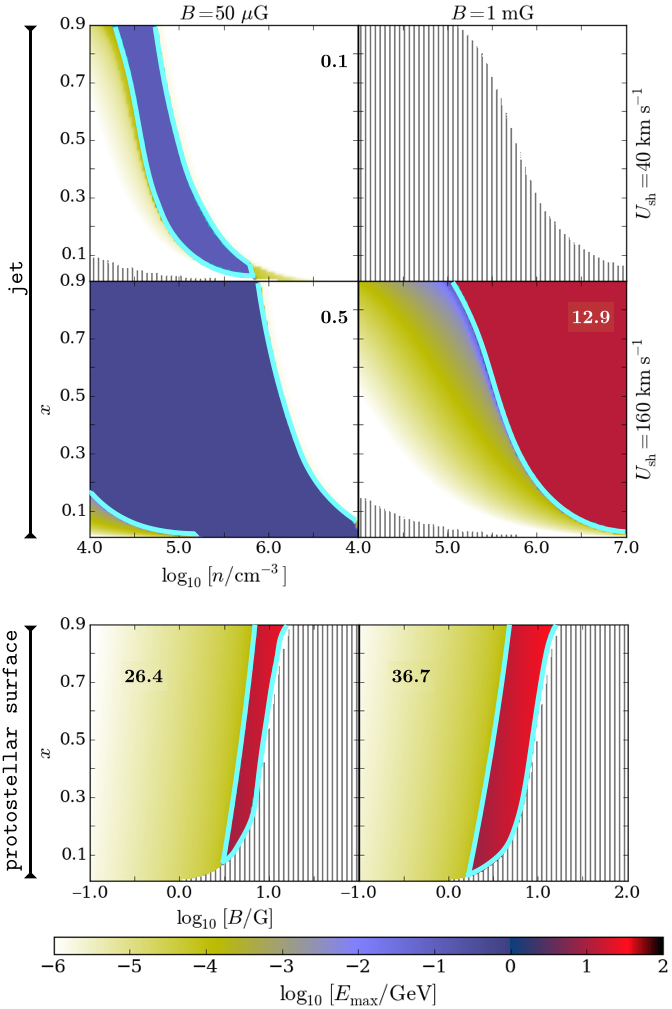}}
\caption{{\em Upper panel.} Case of a parallel shock in jets: ionisation fraction, $x$, versus density of neutrals, $n$, 
for $U_{\rm sh}=40$~and~$160$~km~s$^{-1}$, $B=50~\mu$G~and~1~mG, and $T=10^{4}$~K.
{\em Lower panel.} Case of parallel ({\em left}) and perpendicular ({\em right}) 
shocks on protostellar surfaces: ionisation fraction, $x$, versus magnetic field strength, $B$, for
$U_{\rm sh}=260$~km~s$^{-1}$, $T=9.4\times10^{5}$~K, and $n=1.9\times10^{12}$~cm$^{-3}$.
Colour maps show values of $E_{\rm max}$ for protons and
{\em cyan} contours delimit the regions where $E_{\rm max}$ reaches its maximum constant value shown in GeV in each subplot.
{\em Vertically hatched regions} refer to combinations of parameters corresponding to 
strong wave damping ($\mathscr{R}<1$).}
\label{nxT4}
\end{center}
\end{figure}

For jet shocks we perform a parametric study 
using the values in the second row of Table~\ref{paramspace}, 
assuming a parallel shock, 
$\eta=10^{-5}$, and $T=10^{4}$~K. We also consider 
$\kappa_{\rm u}=\kappa_{\rm B}$, which is the most favourable circumstance for
accelerating particles (Drury et al.~\cite{d83}) and we compute $E_{\rm max}$ for $R_{\rm sh}=100$~AU and $R_{\perp}=10$~AU.
The upper panel of Fig.~\ref{nxT4}
shows the maximum energy which a shock-accelerated proton can reach.
By increasing both $U_{\rm sh}$ and $B$, $E_{\rm max}$
attains higher values up to about 13~GeV for protons. 
Once the combination of parameters satisfies the condition $\mathscr{R}>1$,
$E_{\rm max}$ rapidly reaches a constant value, encompassed by the cyan contours in each subplot. 
In fact, the maximum energy is 
controlled by
$E_{\rm esc,d}$ that is independent of both $n$ and $x$.
Supposing the magnetic field to have a strong toroidal component, we repeat the calculation for the case of a perpendicular shock finding
that $E_{\rm max}$ decreases by a factor of about 1.3, 
since $E_{\rm esc,d}^{\perp}/E_{\rm esc,d}^{\parallel}\propto (r+1)/(2r)$.\footnote{Superscripts $\perp$ and $\parallel$ refer
to perpendicular and parallel shocks, respectively.}

For shocks on protostellar surfaces, we use values in the third row of Table~\ref{paramspace}, varying $x$ and $B$.
Assuming $\eta=10^{-5}$, $k_{\rm u}=1$, and a parallel shock, we find values of $E_{\rm max}$ for protons 
up to about 26~GeV for $B\sim3-10$~G  
(see lower panel of Fig.~\ref{nxT4}),
which are
comparable with magnetic field intensities computed by e.g. Garcia et al.~(\cite{gf01}). 
Because of high temperatures, Coulomb losses are dominant and $E_{\rm max}$ is constrained by $E_{\rm loss}$. Thus, 
for a perpendicular shock, $E_{\rm max}$ is a factor of about 1.3 larger, since $E_{\rm loss}^{\perp}/E_{\rm loss}^{\parallel}\propto (r+1)/r$.
Hatched areas in both panels of Fig.~\ref{nxT4} show regions where acceleration is not possible because of strong wave damping.

Electrons can be accelerated as well, but generally $E_{\rm max}$ for electrons is much smaller than $E_{\rm max}$ for protons because of wave damping and stronger energy losses.
As an instance, for $U_{\rm sh}=160$~km~s$^{-1}$ and
$B=1$~mG, $E_{\rm max}\sim300$~MeV for a narrow range of density and ionisation fraction ($n\gtrsim3\times10^{6}$~cm$^{-3}$, $x\gtrsim0.6$). 
For lower values of $B$ and $U_{\rm sh}$, $E_{\rm max}\lesssim50$~MeV.
We also find that electron acceleration is not triggered by protostellar surface shocks.

\section{Discussion and conclusions}\label{conclusions}
We investigated the possibility of accelerating particles within a protostellar source by means of shock processes
through the diffusive shock acceleration mechanism. 
We focused our attention on the effectiveness of shocks in
envelopes, on protostellar surfaces, and in jets. We concluded that:
\begin{itemize}
\item[$(i)$] In envelopes, $x$ and $U_{\rm sh}$ are too small, preventing
any particle acceleration. Besides $B$ is as large as to yield sub-Alfv\'enic shocks.
\item[$(ii)$] Jet shocks are possible accelerators of particles that can be easily boosted up to relativistic energies. The acceleration
is more efficient for protons which can reach up to about 13 or 10~GeV (for parallel or perpendicular shocks, 
respectively), 
while electrons attain at most about 300~MeV because of wave damping and
energy losses.
\item[$(iii)$] Protostellar surface shocks can accelerate protons up to about 26 or 37~GeV (for parallel or perpendicular shocks, 
respectively). Electrons cannot be accelerated mainly because of large magnetic field strengths leading to 
synchrotron losses.
\end{itemize}

The set of conditions that has to be fulfilled is highly non-linear: small variations in one or more parameters 
($B$, $x$, $n$, $T$, $U_{\rm sh}$, $\eta$, $k_{\rm u}$, $k_{\rm d}$) can 
make the acceleration process inefficient. As a consequence, since young protostars are highly dynamic systems, 
particle acceleration can be a very intermittent process.
In a following paper we will discuss in detail other possible acceleration mechanisms as well as 
the effect of variations in the parameter set and the distance of the jet shocks from the protostar, 
including departures from Bohm-type diffusion
with a view to studying the propagation of
high-energy particles in the protostellar environment in order to explain the available observations.

\begin{acknowledgements} 
The authors thank Elena Amato, Francesca Bacciotti, Sylvie Cabrit, Claudio Codella, Daniele Galli, and Linda Podio
for valuable discussions about jets in protostars 
and shock properties.
We acknowledge the financial support of the Agence National
pour la Recherche (ANR) through the COSMIS project. 
This work has been carried out thanks to the support of the OCEVU Labex (ANR-11-LABX-0060) 
and the A*MIDEX project (ANR-11-IDEX-0001-02) funded by the ``Investissements d'Avenir'' French government programme managed by the ANR.
MP and AM also acknowledge the support of the CNRS-INAF PICS project ``Pulsar wind nebulae,
supernova remnants and the origin of cosmic rays''.
\end{acknowledgements}

\end{document}